\documentclass{article}
\usepackage{spconf,amsmath,graphicx,hyperref}
\usepackage{booktabs}
\usepackage{multirow}
\usepackage{makecell}

\newcommand{\ourresult}{$\epsilon$ar-VAE}

\title{Back to Ear: Perceptually Driven High Fidelity Music Reconstruction}
\name{Kangdi Wang$^{*1}$ \qquad Zhiyue Wu$^{*2}$ \qquad Dinghao Zhou$^{3}$ \qquad Rui Lin$^{4}$ \qquad Junyu Dai$^{\S\dagger}$ \qquad Tao Jiang$^{\dagger}$}

\address{{$\epsilon$ar-LAB} \qquad {initi-AI Ltd}}

\contribution[*]{Core Contribution}
\contribution[\S]{Corresponding Author}
\contribution[\dagger]{Project Lead}

\begin{document}
\topmargin=0mm
\ninept
\maketitle
\begin{abstract}
Variational Autoencoders (VAEs) are essential for large-scale audio tasks like diffusion-based generation. However, existing open-source models often neglect auditory perceptual aspects during training, leading to weaknesses in phase accuracy and stereophonic spatial representation. To address these challenges, we propose {\ourresult}, an open-source music signal reconstruction model that rethinks and optimizes the VAE training paradigm. Our contributions are threefold: (i) A K-weighting perceptual filter applied prior to loss calculation to align the objective with auditory perception. (ii) Two novel phase losses: a Correlation Loss for stereo coherence, and a Phase Loss using its derivatives—Instantaneous Frequency and Group Delay—for precision. (iii) A new spectral supervision paradigm where magnitude is supervised by all four MSLR (Mid/Side/Left/Right) components, while phase is supervised only by the LR components. Experiments show {\ourresult} at 44.1kHz substantially outperforms leading open-source models across diverse metrics, showing particular strength in reconstructing high-frequency harmonics and the spatial characteristics.
\end{abstract}

\section{Introduction}
\label{sec:intro}

Achieving perfect, perceptually lossless reconstruction of complex audio signals like music remains a central challenge in audio engineering and machine learning. High-fidelity audio Variational Autoencoders (VAEs) \cite{evans2024stableaudioopen} are foundational reconstructive components for many downstream tasks, which fundamentally differs from that of traditional generative VAEs like MusicVAE \cite{roberts2019hierarchicallatentvectormodel}. While the latter prioritizes the generation of semantically authentic novel content, the former aims to compress and decompress the original signal losslessly. To achieve this, the model prioritize perceptually significant details and discarding imperceptible information. This process relies heavily on psychoacoustic principles, such as utilizing perceptual weighting curves like A-weighting or K-weighting, to model the frequency-dependent sensitivity of human hearing. However, modern audio VAE models fail to integrate such fine-grained perceptual weighting strategies into their training paradigms.

Furthermore, the reconstruction of high-quality music requires the accurate modelling of both phase and spatial information. Spatial information, often parameterized by the Mid/Side (M/S) decomposition, is critical for accurately rendering the stereo image. Concurrently, audio transients and clarity are determined not by the absolute phase of Short-time Fourier Transform (STFT) bins, but by their partial derivatives: Instantaneous Frequency (IF) across time and Group Delay (GD) across frequency. However, existing open-source models lack effective mechanisms to supervise these critical phase derivatives and fail to fully leverage the M/S representation for spatial reconstruction, which leads to audible artifacts, such as transient smearing and an inaccurate stereo image, limiting their use in professional applications.

\begin{figure}[h]
    \centering
    \includegraphics[width=0.45\textwidth]{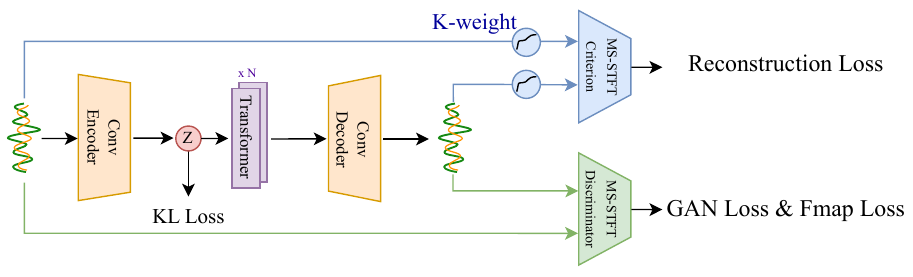}

    \caption{Architecture of {\ourresult}}
    \label{fig:ear-vae}
\end{figure}

To address these shortcomings, we introduce {\ourresult}, an open-source VAE model optimized for high-fidelity music reconstruction. Our model incorporates a K-weighting perceptual filter, which we demonstrate is psychoacoustically better suited for music signals than A-weighting. To ensure phase coherence, we propose novel loss functions that implicitly optimize the phase performance by supervising its derivatives (IF \& GD). Additionally, we apply the reconstruction loss with a new Mid/Side/Left/Right (MSLR) weighting scheme to maximize the preservation of both spatial and spectral details. Through these targeted designs, {\ourresult} achieves state-of-the-art reconstruction performance across multiple objective evaluations, setting a new benchmark for open-source high-fidelity audio VAEs.

We summarize our contributions as follows: Firstly, we analyse and integrate the K-weighting filter into the VAE training pipeline, aligning the reconstruction objective with psychoacoustics of music perception, in contrast to the commonly A-weighting. Secondly, we propose novel phase-aware loss functions that supervise phase derivatives to implicitly model critical phase differences, thereby enhancing transient clarity and phase coherence. Thirdly, we introduce a novel supervision strategy that separately constrains magnitude and phase, employing all MSLR components for magnitude reconstruction while using only LR to ensure phase coherence.

Now the codes, model weights and examples are available now\footnote{\url{https://eps-acoustic-revolution-lab.github.io/EAR_VAE/}}.

\section{Related Work}

While traditional generative VAEs \cite{vae} utilize a Kullback-Leibler (KL) divergence loss to enforce a continuous Gaussian prior for generation, the reconstruction task prioritizes the model's ability to faithfully compress and restore signals, which have shifted towards discrete quantized representations. This approach, pioneered by VQ-VAE \cite{oord2018neuraldiscreterepresentationlearning} and now standard in neural audio codecs like EnCodec \cite{défossez2022highfidelityneuralaudio} and DAC \cite{kumar2023highfidelityaudiocompressionimproved}, excels at achieving high compression ratios. To enhance the perceptual quality of the decoded audio, these frameworks often incorporate a powerful adversarial component, leveraging discriminators from vocoders like MelGAN \cite{kumar2019melgangenerativeadversarialnetworks} and HiFi-GAN \cite{kong2020hifigangenerativeadversarialnetworks}. Despite their success, the inherent information loss from quantizing remains a fundamental limitation, as subtle details crucial for reconstruction can be discarded at the bottleneck.

In contrast, the approach revisiting continuous latent representations offers a potentially higher-fidelity pathway for reconstruction. The VAE model from Stable-Audio-Open (SAO) \cite{evans2024stableaudioopen} stands as a prominent example of this approach, employing a VAE-GAN framework with adversarial loss and a down-weighted KL divergence to learn a continuous representation at a high compression rate.

\section{\ourresult}

Our model, inspired by the VAE architecture of SAO, is a partially convolutional VAE complemented by transformer-based bottleneck layers, trained with a composite adversarial objective. As shown in figure \ref{fig:ear-vae}, the overall architecture consists of a traditional VAE generator and a powerful time-frequency domain discriminator. The generator encodes the input waveform into a latent representation and then decodes it back into a waveform, while the discriminator distinguishes audios between real and reconstructed version, thereby guiding the generator to produce higher-fidelity output.

\subsection{Generator}

Our generator employs an encoder-decoder architecture featuring several key designs optimized for music reconstruction. The encoder utilizes a series of strided convolutional blocks with the SnakeBeta activation function \cite{lee2023bigvganuniversalneuralvocoder}, which outperforms alternatives such as ELU in our experiment. The decoder is designed asymmetrically: it mirrors the encoder's convolutional structure using transposed convolutions for upsampling but also incorporates a powerful transformer module with RoPE position embeddings \cite{su2023roformerenhancedtransformerrotary} on the decoding path. This asymmetric design delegates local feature extraction to the efficient encoder, while the decoder's transformers model global dependencies, yielding superior performance over symmetric architectures like Mimi \cite{défossez2024moshispeechtextfoundationmodel}. We selected transposed convolutions rather than upsampling-plus-convolution because the former preserves greater signal energy and perceptual loudness, which is more significant than slightly higher high-frequency clarity offered by the latter. Finally, all convolutional layers are weight-normalized \cite{evans2024stableaudioopen} for training stability.

\subsection{Discriminator}
For adversarial training, we employ a Multi-Resolution STFT Discriminator (MR-STFTD) as MSD, inspired by Encodec \cite{défossez2022highfidelityneuralaudio}. This approach assesses the signal across various STFT resolutions, enabling it to detect a wide range of artifacts from coarse spectral errors to fine-grained phase inconsistencies. Notably, we omit the Multi-Period Discriminator (MPD). Our experiments show that MPD introduces spatial positioning artifacts in the stereo field, which we attribute to its fixed periodic analysis being ill-suited for the complex, inconstant rhythms of music. In contrast, a single MSD provides robust supervision without introducing such spatial distortions.

\begin{figure}
    \centering
    \includegraphics[width=0.7\linewidth]{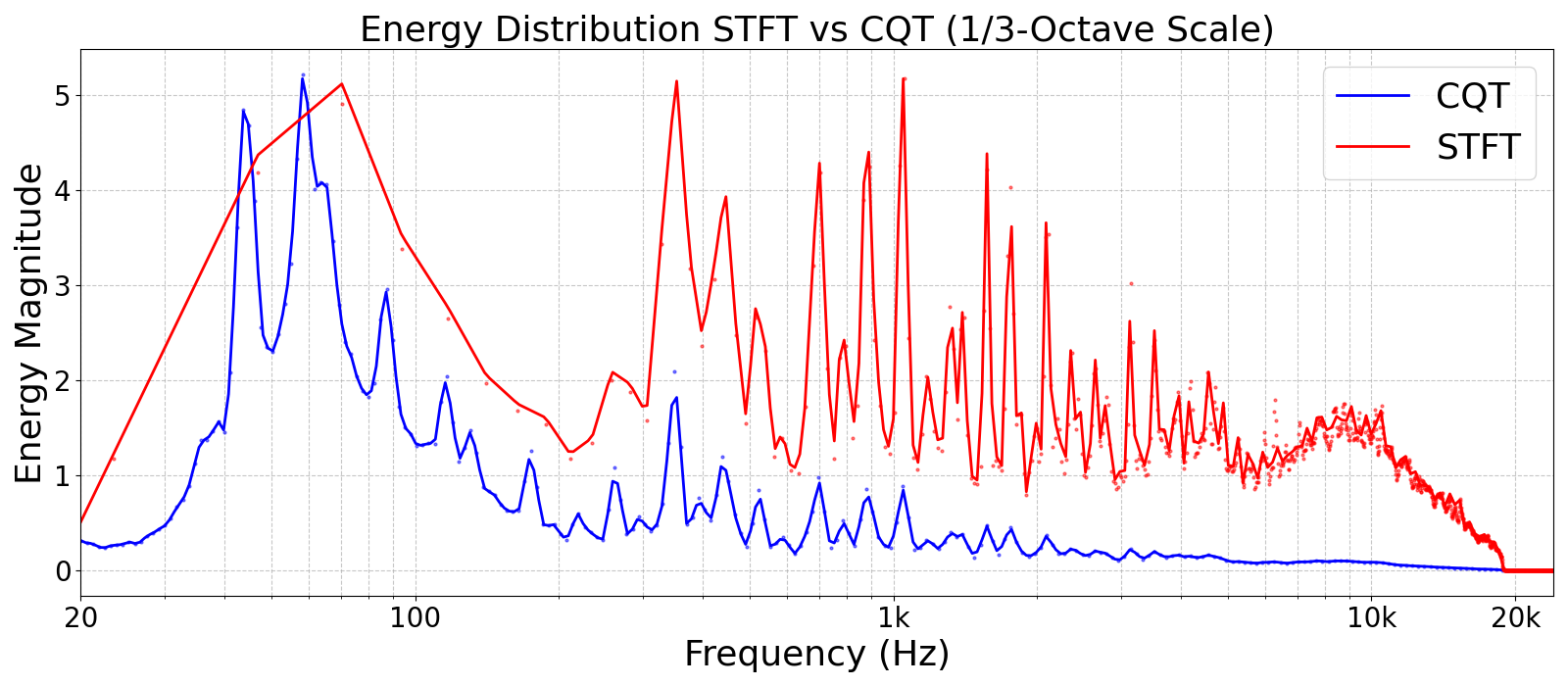}
    \caption{Energy-based frequency response comparison between CQT and STFT of a musical excerpt.}
    \label{fig: STFT vs. CQT}
\end{figure}

While some works like BigVGAN \cite{lee2023bigvganuniversalneuralvocoder} apply the complementary MSD in Constant-Q Transform (CQT) representation, we find this approach degrades the VAE's ability to represent music. As shown in figure \ref{fig: STFT vs. CQT}, CQT overemphasizes low-to-mid frequency melodic features, leading to the lack of high-frequency harmonics, whereas the STFT provides a more balanced and suitable frequency response for our model.

\subsection{Loss functions}

\phantomsection\label{subsub: ms-mag-loss}
\textbf{Multi-Scale Log-Magnitude Loss}\quad To guide the generator's spectral amplitude reconstruction, we adopt the multiscale STFT loss formulation from EnCodec \cite{défossez2022highfidelityneuralaudio}. This loss, denoted as $\mathcal{L}_{\mathrm{stft-mag}}$, computes the L1 distance between the logarithmic magnitudes of the predicted and target spectrograms over a set of different STFT resolutions, effectively capturing both coarse harmonic structures and fine temporal details.

\phantomsection\label{subsub: fmap-loss}
\textbf{Feature-Map Loss}\quad We also employ the feature-matching loss from EnCodec \cite{défossez2022highfidelityneuralaudio} to enhance perceptual quality. This loss, $\mathcal{L}_{\mathrm{fmap}}$, minimizes the L1 distance between the intermediate feature maps extracted from the discriminator's layers for the ground-truth and generated audio.

\phantomsection\label{subsub: GAN-loss}
\textbf{Adversarial Loss}\quad We employ the least-squares GAN objective from BigVGAN \cite{lee2023bigvganuniversalneuralvocoder} for adversarial training, defining the generator loss $\mathcal{L}_{\text{Adv}}(G)$ and the discriminator loss $\mathcal{L}_{\text{Adv}}(D)$.

\phantomsection\label{subsub: kl-loss}
\textbf{KL Loss}\quad We regularize the latent space using a Kullback-Leibler (KL) divergence loss, which aligns the posterior distribution $q(z|x)$ with a standard normal prior $\mathcal{N}(\mathbf{0}, \mathbf{I})$ to promote a continuous and well-structured representation.

\phantomsection\label{subsub: corr-loss}
\textbf{Correlation Loss}\quad Inspired by music production metrics, our \textit{Correlation Loss} directly penalizes phase deviations between the ground truth spectrogram $\mathbf{S}$ and its reconstruction $\hat{\mathbf{S}}$ is defined as: 
\begin{equation}
    \mathcal{L}_{\mathrm{corr}} = 1 - \sum_{}^{} \mathrm{Re}\left( \frac{\hat{S} \overline{S}}{\left| \hat{S} \right| \left| S \right| + \varepsilon} \right),
    \label{eq:phase_corr_loss}
\end{equation}
where the term within the summation normalizes the cross-power spectrum, simplifying to the cosine of the phase difference between the signals, $\cos(\phi_{\hat{S}} - \phi_{S})$. Minimizing this loss thus encourages perfect phase coherence.

\phantomsection\label{subsub: phase-loss}
\textbf{Phase Loss}\quad Phase instability often introduces ``electrical buzz" artifacts. To mitigate this, our \textit{Phase Loss} constrains the phase's first-order partial derivatives: Instantaneous Frequency (IF) and Group Delay (GD). These derivatives are computed via finite differences on the phase $\phi=\arg(S)$ and $\hat{\phi}=\arg(\hat{S})$:
\begin{equation}
\left\{
\begin{aligned}
    \mathrm{IF}(\phi)_t &= \phi_{t+1} - \phi_t, \\
    \mathrm{GD}(\phi)_f &= -(\phi_{f+1} - \phi_f),
\end{aligned}
\right.
\label{eq:if_gd_definition}
\end{equation}
where the subscripts $t$ and $f$ denote the time and frequency dimensions, respectively. All phase differences are computed modulo $2\pi$ to resolve the discontinuity at the $\pm\pi$ boundary. The total loss penalizes the $L1$ norm between the ground-truth and estimated derivatives, to optimize phase coherence in a more stable and perceptually relevant manner than direct phase supervision.

\begin{figure}
    \centering
    \includegraphics[width=0.6\linewidth]{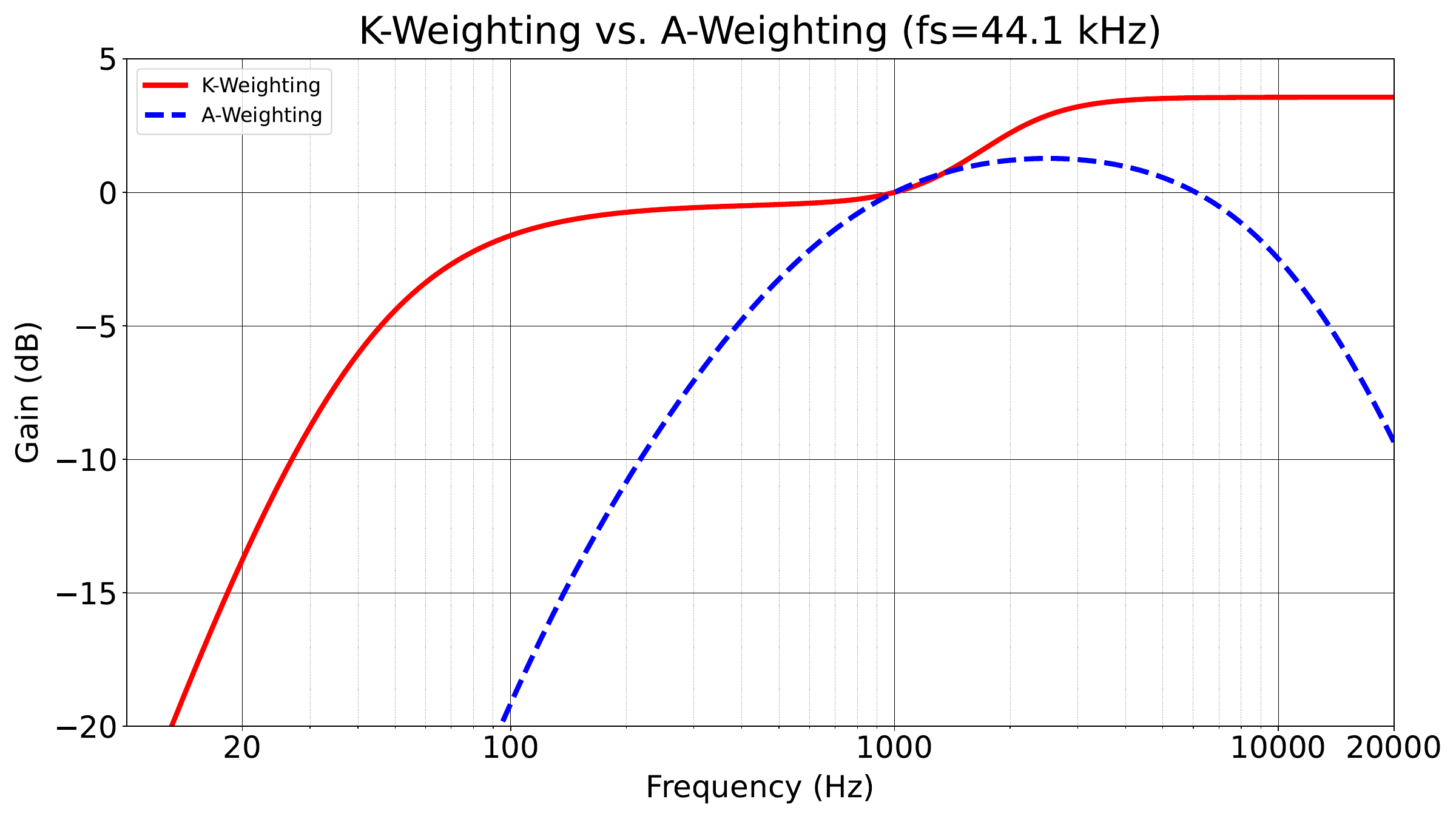}
    \caption{K-weighting vs. A-weighting Curve}
    \label{fig:kw_curve}
\end{figure}

\phantomsection\label{subsub: k-weight}
\textbf{K-Weighting Curve}\quad K-weighting, originating from ITU-R BS.1770 \cite{ITU-R_BS1770-5} loudness measurement standards, which is widely applied in music production, is designed to approximate the frequency-dependent sensitivity of the human ear in \ref{fig:kw_curve}. This cascaded filter accentuates mid and high frequency bands where human hearing is most sensitive, while attenuating lower frequencies. By pre-filtering the signal with $H_{\mathrm{K}}(z)$, we ensure that reconstruction losses are evaluated in a perceptually relevant domain.

\phantomsection\label{subsub: MSLR}
\textbf{M-S-L-R Split}\quad Inspired from the music mixing process and SAO \cite{evans2024stableaudioopen}, we also apply the stereophonic split in two different representations, Left-Right ($S_L$, $S_R$) and Mid-Side ($S_M$, $S_S$), where mid is defined as $(S_L + S_R) / 2$, side $(S_L - S_R) / 2$. For different supervision attributes, we take certain combinations of the generator and the discriminator separately:

\phantomsection\label{subsub: total-loss}
\textbf{Total Loss}\quad The complete training objective consists of two parts: one for the generator $\mathcal{L}_{G}$ and one for the discriminator $\mathcal{L}_{D}$. These are optimized in alternating steps:
\begin{equation}
\left\{
\begin{aligned}
    \mathcal{L}_{G} &= \begin{aligned}[t]
        &\lambda_{\mathrm{stft-mag}} \mathcal{L}_{\mathrm{stft-mag}} + \lambda_{\mathrm{corr}} \mathcal{L}_{\mathrm{corr}} + \lambda_{\mathrm{phase}} \mathcal{L}_{\mathrm{phase}}\\
        & + \lambda_{\mathrm{fmap}} \mathcal{L}_{\mathrm{fmap}} + \lambda_{\mathrm{adv}} \mathcal{L}_{\mathrm{Adv}}(G; D) + \lambda_{\mathrm{KL}}\mathcal{L}_{\mathrm{KL}},
    \end{aligned} \label{eq:g_loss_aligned} \\
    \mathcal{L}_{D} &= \mathcal{L}_{\mathrm{Adv}}(D; G),
\end{aligned}
\right.
\end{equation}

where $\lambda_{\{\cdot\}}$ are hyperparameters that control the relative importance of each loss component.

\begin{table*}[h]
\centering
\caption{Results on MuChin and In-house validation split (side-by-side comparison).}
\label{tab:final_combined_results}

\setlength{\tabcolsep}{3.5pt}

\resizebox{\textwidth}{!}{
\begin{tabular}{llr|cccccc|cccccc}
\toprule
\multirow{3}{*}{\textbf{Model}} & 
\multirow{3}{*}{\begin{tabular}{@{}c@{}}\textbf{Channels}/\\\textbf{Rate (Hz)}\end{tabular}} & 
\multirow{3}{*}{\begin{tabular}{@{}c@{}}\textbf{Latent}\\\textbf{Rate}\end{tabular}} &
\multicolumn{6}{c|}{\textbf{MuChin}} & \multicolumn{6}{c}{\textbf{In-house validation}} \\
\cmidrule(lr){4-9} \cmidrule(lr){10-15}
& & &
\makecell{Mel \\ dist}$\downarrow$ & 
\makecell{STFT \\ dist}$\downarrow$ & 
ICPC$\uparrow$ & CCPC$\uparrow$ & 
SI-SDR$\uparrow$ & \makecell{dbTP \\ dist}$\downarrow$ &
\makecell{Mel \\ dist}$\downarrow$ & 
\makecell{STFT \\ dist}$\downarrow$ & 
ICPC$\uparrow$ & CCPC$\uparrow$ & 
SI-SDR$\uparrow$ & \makecell{dbTP \\ dist}$\downarrow$ \\
\midrule

DAC & 1/44.1k & 86Hz 
& 0.71 & 1.33 & 94.25\% & 90.69\% & 6.14 & 0.29
& 0.67 & 1.21 & \underline{94.49\%} & 90.47\% & 6.68 & 0.35 \\

Encodec & 2/48k & 50Hz 
& 0.84 & 1.57 & 89.89\% & 89.34\% & 3.64 & \underline{0.10}
& 0.80 & 1.49 & 90.07\% & 89.42\% & 3.99 & 0.14 \\

AGC & 2/48k & 100Hz 
& 0.71 & 1.46 & \underline{94.70\%} & \underline{94.96\%} & \underline{7.58} & 0.27
& 0.65 & 1.39 & 94.16\% & \underline{94.66\%} & \underline{8.21} & \underline{0.09} \\

SAO & 2/44.1k & 21.5Hz 
& 0.75 & 1.64 & 90.70\% & 91.41\% & 4.62 & 0.29
& 0.64 & 1.34 & 90.37\% & 91.12\% & 5.23 & 0.36 \\

\ourresult & 2/44.1k & 43Hz 
& \textbf{0.55} & \textbf{1.17} & \textbf{96.78\%} & \textbf{96.81\%} & \textbf{9.99} & \textbf{0.05}
& \textbf{0.55} & \textbf{1.12} & \textbf{96.66\%} & \textbf{96.52\%} & \textbf{11.00} & \textbf{0.05} \\

\bottomrule
\end{tabular}
}
\end{table*}

\section{Experiments and Results}

\subsection{Datasets}
Our model is trained on a combination of large-scale public datasets and a high-quality, proprietary in-house dataset. The training process is conducted in two stages: pre-training and continue-training. First, we use a diverse mix of public data including FSD50K \cite{fonseca2022FSD50K}, FMA \cite{def2017fma_dataset}, and DISCO-10M \cite{lanzendörfer2023disco10mlargescalemusicdataset}. Subsequently, for the continue-training stage,  the model is trained on our in-house dataset of approximately 10,000 hours of professionally produced music.

\subsection{Data Pipeline}

To ensure data quality, we designed a two-stage filtering pipeline, applied progressively. All datasets undergo Stage 1 for format standardization, while only our in-house data is subjected to the full pipeline.

\textbf{Stage 1: Format and Loudness Standardization}\quad 
All the files are formatted to 44.1 kHz stereo, with files natively sampled below this rate being discarded. Next, we filter based on perceived loudness. Using the LUFS-I metric \cite{ebu-r128}, we retain only tracks with an integrated loudness between -22 and -5 LUFS, removing acoustically extreme examples.

\textbf{Stage 2: True Peak Filtering}\quad 
For our in-house data, we filter true peak levels less than +1dB to handle signal clipping. Different from the any-clip-rejection strategy from Encodec \cite{kumar2023highfidelityaudiocompressionimproved}, this lenient criterion is deliberately chosen to accommodate the moderate, intentional clipping common in modern music mastering.

\begin{figure*}
  \centering
  \includegraphics[width=0.8\textwidth]{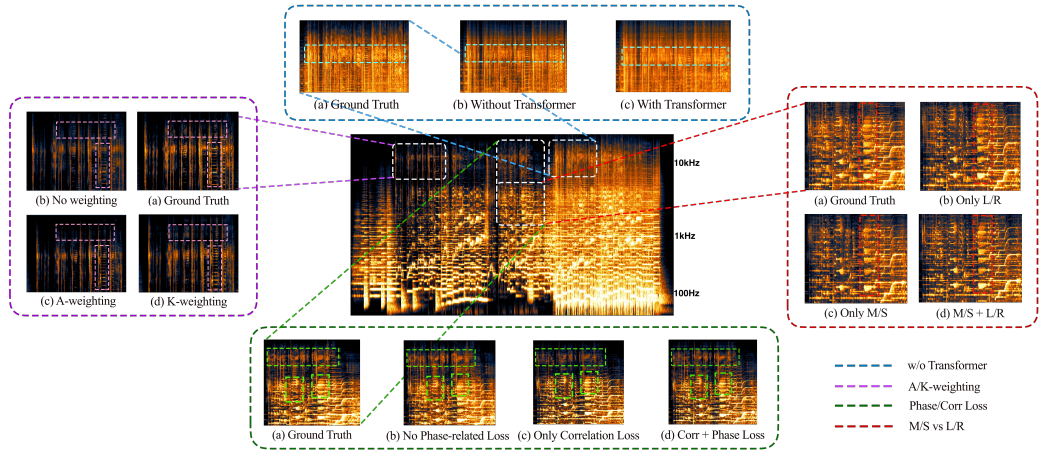}
  \caption{All Ablation Study}
  \label{fig: all_ablation_study}
\end{figure*}

\subsection{Training Details}
The {\ourresult} achieves a 1024x compression rate via a 5-layer convolutional encoder/decoder structure with strides $[2, 4, 4, 4, 8]$ and a 128-dimensional latent space. The decoder is augmented with two transformer layers using RoPE. The multiscale discriminator uses STFT window sizes of $[2048, 1024, 512, 256, 128]$. The full model contains 141M parameters.

We train all models on 8 A100 GPUs using the AdamW optimizer with a learning rate of $3 \times 10^{-4}$, $\beta_1=0.5$, and $\beta_2=0.9$. The loss weights are set as follows: $\lambda_{\mathrm{stft-mag}}=50$, $\lambda_{\mathrm{corr}}=10$, $\lambda_{\mathrm{phase}}=10$, $\lambda_{\mathrm{fmap}}=20$, $\lambda_{\mathrm{adv}}=1$, and $\lambda_{\mathrm{KL}}=10^{-6}$ for reconstruction.

We train the model for a total of 2M steps over three distinct phases:

\textbf{Phase 1: Warm-up (10k steps)}\quad We train only the generator using the STFT and KL losses, accompanied by a linear learning rate warm-up.

\textbf{Phase 2: Pre-train (1M steps)}\quad On the public datasets, we activate all loss components and train the generator and discriminator alternately. The learning rate is halved at 200k, 400k, and 600k steps.

\textbf{Phase 3: Continue-train (1M steps)}\quad We continue training on our in-house dataset using the same full loss configuration and alternate training scheme as in Phase 2, with a constant learning rate.

\subsection{Results}
\subsubsection{Novel Evaluation Metrics}
\label{subsub: ICPC&CCPC}
To evaluate phase accuracy, we introduce two custom metrics, Individual Channel Phase Coherence and Cross Channel Phase Coherence.

\textbf{Individual Channel Phase Coherence (ICPC)}\quad 
ICPC quantifies the stability of phase errors within each channel. It is derived by first computing the phase error $\Delta\phi(f, t)$ at each time-frequency bin, with magnitude-based weighting applied to mitigate the influence of noise. For each time frame, a coherence score $\mathcal{C}_t$ is calculated as the mean resultant length of these weighted phase error phasors. The final ICPC score is the energy-weighted average of these per-frame scores, where the energy $E_t$ for each frame is the sum of its corresponding weights.
\begin{equation}
    \text{ICPC} = \frac{\sum_{t} \mathcal{C}_t \cdot E_t}{\sum_{t} E_t + \epsilon}.
    \label{eq:icpc}
\end{equation}

\textbf{Cross Channel Phase Coherence (CCPC)}\quad 
CCPC extends this concept to stereo signals by measuring the preservation of the Inter-channel Phase Difference (IPD). Its calculation follows the same principle as ICPC, but is based on the error in the IPD instead of the single-channel phase error. The final score is similarly the energy-weighted average across time, where the frame energy $E_{\text{inter}, t}$ is the sum of the corresponding inter-channel weights.
\begin{equation}
    \text{CCPC} = \frac{\sum_{t} \mathcal{C}_{\text{inter}, t} \cdot E_{\text{inter}, t}}{\sum_{t} E_{\text{inter}, t} + \epsilon}.
    \label{eq:ccpc}
\end{equation}

\subsubsection{Evaluation Setting}
We evaluate performance using several objective metrics, primarily sourced from the auraloss library \cite{steinmetz2020auraloss}: Multi-Scale STFT (MS-STFT) distance, Multi-Scale Mel (MS-Mel) distance, and SI-SDR. The multiscale metrics were configured with FFT sizes of $[4096, 2048, 1024, 512, 256, 128]$ and a hop size of one-quarter the window size. Additionally, we measure the True Peak loudness difference (dbTP) using FFmpeg \cite{tomar2006converting} and the designed metrics in section \ref{subsub: ICPC&CCPC}.

We compare {\ourresult} against several leading audio reconstruction models: EnCodec, DAC, AudioGen (AGC), and Stable-Audio-Open (SAO). The evaluation is performed on the reconstruction of test sets from the MuChin \cite{wang2024muchin} and our in-house validation datasets. Detailed results are presented in table \ref{tab:final_combined_results}.

\subsection{Ablation Study}

Our design choices are validated through a series of ablation studies, with qualitative results visualized in figure \ref{fig: all_ablation_study}, which compares reconstructed spectrograms from various model configurations against the ground truth.

\textbf{Impact of Architectural Components}\quad 
The top panel of figure \ref{fig: all_ablation_study} highlights the role of the transformer layers. Without them, the model fails to reconstruct fine-grained harmonic structures above 10 kHz, confirming that the transformer's self-attention is crucial for modelling long-range frequency dependencies, complementing the convolutional layers' feature extraction.

\textbf{Impact of Perceptual and Phase-related Losses}\quad 
The left and bottom panels demonstrate the effects of our proposed loss functions. As shown in the left panel, removing the K-weighting pre-filter results in a suboptimal reconstruction, particularly in the critical mid-to-high frequency bands, while the A-weighting curve improperly attenuates high frequencies. The bottom panel illustrates that removing the phase-related losses leads to a loss of clarity and the introduction of audible "current-like" noise. Specifically, the Phase Loss ensures local phase smoothness, while the Correlation Loss contributes to enhances spectral coherence, particularly for polyphonic elements.

\textbf{Impact of Stereo and Spectral Representation}\quad 
Our ablation results reveal a key principle in stereo supervision. For magnitude, supervision over all four components—left, right, mid, and side—provides a more complete guidance signal for stereo reconstruction. In contrast, phase supervision should be restricted to the pure left/right mode. Incorporating mid/side components into phase losses distorts the physically meaningful Inter-aural Phase Difference (IPD) cues, thereby introducing spatial artifacts.

\section{Conclusion}
In this paper, we present {\ourresult}, a variational autoencoder that sets a new state-of-the-art for high-fidelity music reconstruction. Our ablation studies confirm that the carefully designed components, including novel perceptual and phase-based losses, contribute significantly to the superior performance. Furthermore, We believe that addressing the model's tendency to attenuate subtle spatial effects by exploring targeted loss functions remains a promising research direction and tackling these challenges will unlock the next generation of controllable and realistic generative music models.
 
\bibliographystyle{IEEEbib}
\bibliography{refs}

@misc{evans2024stableaudioopen,
      title={Stable Audio Open}, 
      author={Zach Evans and Julian D. Parker and CJ Carr and Zack Zukowski and Josiah Taylor and Jordi Pons},
      year={2024},
      eprint={2407.14358},
      archivePrefix={arXiv},
      primaryClass={cs.SD},
      url={https://arxiv.org/abs/2407.14358}, 
}

@article{vae,
  title={Auto-encoding variational bayes},
  author={Kingma, Diederik P and Welling, Max},
  journal={arXiv preprint arXiv:1312.6114},
  year={2013}
}

@misc{oord2018neuraldiscreterepresentationlearning,
      title={Neural Discrete Representation Learning}, 
      author={Aaron van den Oord and Oriol Vinyals and Koray Kavukcuoglu},
      year={2018},
      eprint={1711.00937},
      archivePrefix={arXiv},
      primaryClass={cs.LG},
      url={https://arxiv.org/abs/1711.00937}, 
}

@misc{roberts2019hierarchicallatentvectormodel,
      title={A Hierarchical Latent Vector Model for Learning Long-Term Structure in Music}, 
      author={Adam Roberts and Jesse Engel and Colin Raffel and Curtis Hawthorne and Douglas Eck},
      year={2019},
      eprint={1803.05428},
      archivePrefix={arXiv},
      primaryClass={cs.LG},
      url={https://arxiv.org/abs/1803.05428}, 
}

@misc{kumar2019melgangenerativeadversarialnetworks,
      title={MelGAN: Generative Adversarial Networks for Conditional Waveform Synthesis}, 
      author={Kundan Kumar and Rithesh Kumar and Thibault de Boissiere and Lucas Gestin and Wei Zhen Teoh and Jose Sotelo and Alexandre de Brebisson and Yoshua Bengio and Aaron Courville},
      year={2019},
      eprint={1910.06711},
      archivePrefix={arXiv},
      primaryClass={eess.AS},
      url={https://arxiv.org/abs/1910.06711}, 
}

@misc{kong2020hifigangenerativeadversarialnetworks,
      title={HiFi-GAN: Generative Adversarial Networks for Efficient and High Fidelity Speech Synthesis}, 
      author={Jungil Kong and Jaehyeon Kim and Jaekyoung Bae},
      year={2020},
      eprint={2010.05646},
      archivePrefix={arXiv},
      primaryClass={cs.SD},
      url={https://arxiv.org/abs/2010.05646}, 
}

@misc{défossez2022highfidelityneuralaudio,
      title={High Fidelity Neural Audio Compression}, 
      author={Alexandre Défossez and Jade Copet and Gabriel Synnaeve and Yossi Adi},
      year={2022},
      eprint={2210.13438},
      archivePrefix={arXiv},
      primaryClass={eess.AS},
      url={https://arxiv.org/abs/2210.13438}, 
}

@misc{kumar2023highfidelityaudiocompressionimproved,
      title={High-Fidelity Audio Compression with Improved RVQGAN}, 
      author={Rithesh Kumar and Prem Seetharaman and Alejandro Luebs and Ishaan Kumar and Kundan Kumar},
      year={2023},
      eprint={2306.06546},
      archivePrefix={arXiv},
      primaryClass={cs.SD},
      url={https://arxiv.org/abs/2306.06546}, 
}

@misc{défossez2024moshispeechtextfoundationmodel,
      title={Moshi: a speech-text foundation model for real-time dialogue}, 
      author={Alexandre Défossez and Laurent Mazaré and Manu Orsini and Amélie Royer and Patrick Pérez and Hervé Jégou and Edouard Grave and Neil Zeghidour},
      year={2024},
      eprint={2410.00037},
      archivePrefix={arXiv},
      primaryClass={eess.AS},
      url={https://arxiv.org/abs/2410.00037}, 
}

@misc{su2023roformerenhancedtransformerrotary,
      title={RoFormer: Enhanced Transformer with Rotary Position Embedding}, 
      author={Jianlin Su and Yu Lu and Shengfeng Pan and Ahmed Murtadha and Bo Wen and Yunfeng Liu},
      year={2023},
      eprint={2104.09864},
      archivePrefix={arXiv},
      primaryClass={cs.CL},
      url={https://arxiv.org/abs/2104.09864}, 
}

@misc{lee2023bigvganuniversalneuralvocoder,
      title={BigVGAN: A Universal Neural Vocoder with Large-Scale Training}, 
      author={Sang-gil Lee and Wei Ping and Boris Ginsburg and Bryan Catanzaro and Sungroh Yoon},
      year={2023},
      eprint={2206.04658},
      archivePrefix={arXiv},
      primaryClass={cs.SD},
      url={https://arxiv.org/abs/2206.04658}, 
}

@TechReport{ITU-R_BS1770-5,
  author       = {{International Telecommunication Union}},
  title        = {{Recommendation ITU-R BS.1770-5: Algorithms to measure audio programme loudness and true-peak audio level}},
  institution  = {International Telecommunication Union},
  year         = {2023},
    url={https://www.itu.int/dms_pubrec/itu-r/rec/bs/R-REC-BS.1770-5-202311-I!!PDF-E.pdf},
}

@inproceedings{def2017fma_dataset,
  title = {{FMA}: A Dataset for Music Analysis},
  author = {Defferrard, Micha\"el and Benzi, Kirell and Vandergheynst, Pierre and Bresson, Xavier},
  booktitle = {18th International Society for Music Information Retrieval Conference (ISMIR)},
  year = {2017},
  archiveprefix = {arXiv},
  eprint = {1612.01840},
  url = {https://arxiv.org/abs/1612.01840},
}

@article{fonseca2022FSD50K,
  title={{FSD50K}: an open dataset of human-labeled sound events},
  author={Fonseca, Eduardo and Favory, Xavier and Pons, Jordi and Font, Frederic and Serra, Xavier},
  journal={IEEE/ACM Transactions on Audio, Speech, and Language Processing},
  volume={30},
  pages={829--852},
  year={2022},
  publisher={IEEE}
}

@techreport{ebu-r128,
  author       = "{European Broadcasting Union}",
  title        = "{EBU R 128: Loudness normalisation and permitted maximum level of audio signals}",
  institution  = "{EBU}",
  year         = "{2023}",
  type         = "Recommendation",
  address      = "Geneva, Switzerland",
  note         = "Provides practical guidelines for implementing loudness normalization based on ITU-R BS.1770.",
  url          = {https://tech.ebu.ch/docs/r/r128.pdf}
}

@misc{lanzendörfer2023disco10mlargescalemusicdataset,
      title={DISCO-10M: A Large-Scale Music Dataset}, 
      author={Luca A. Lanzendörfer and Florian Grötschla and Emil Funke and Roger Wattenhofer},
      year={2023},
      eprint={2306.13512},
      archivePrefix={arXiv},
      primaryClass={cs.SD},
      url={https://arxiv.org/abs/2306.13512}, 
}

@inproceedings{steinmetz2020auraloss,
    title={auraloss: {A}udio focused loss functions in {PyTorch}},
    author={Steinmetz, Christian J. and Reiss, Joshua D.},
    booktitle={Digital Music Research Network One-day Workshop (DMRN+15)},
    year={2020}
}

@article{tomar2006converting,
  title={Converting video formats with FFmpeg},
  author={Tomar, Suramya},
  journal={Linux Journal},
  volume={2006},
  number={146},
  pages={10},
  year={2006},
  publisher={Belltown Media}
}

@inproceedings{wang2024muchin,
  title     = {MuChin: A Chinese Colloquial Description Benchmark for Evaluating Language Models in the Field of Music},
  author    = {Wang, Zihao and Li, Shuyu and Zhang, Tao and Wang, Qi and Yu, Pengfei and Luo, Jinyang and Liu, Yan and Xi, Ming and Zhang, Kejun},
  booktitle = {Proceedings of the Thirty-Third International Joint Conference on
               Artificial Intelligence},
  pages     = {7771--7779},
  year      = {2024},
  month     = {8},
}

\end{document}